# Effect of Social Media Use on Mental Health during Lockdown in India

*Sweta Swarnam*

*Symbiosis Centre for Information Technology,*

*Symbiosis International (Deemed University), Pune, India*

*Email:sweta.swarnam@associates.scit.edu*

## Abstract

*This research paper studies about the role of social media use and increase the risk factor of mental health during covid 19 or lockdown. Although few studies have been conducted on the role about the effect of social media use on mental health during lockdown and impact on human reactive nature during lockdown. As a rapidly spreading pandemic, a biomedical disease has serious physical and tremendous mental health implications. An occupational community of internal migrant workers is one of the most vulnerable, but neglected, and is likely to develop psychological ill-effects due to COVID-19's double whammy impact. Mental health is a crucial aspect that needs to be addressed during this lock-down as all modes of communication revolve around the virus. There are many difficulties with the unprecedented changes that have occurred so quickly due to the pandemic and stay-at - home confinement to achieve social distance and mitigate the risk of infection. These include impaired health, well-being, and sleep as a result of daily routine disruption, anxiety, worry, isolation, greater stress on family and work, and excessive screen time. An essential part of our overall health and well-being is mental and emotional health. An important skill is managing emotions and maintaining emotional balance. It helps you face challenges and stress when you manage your emotional health. Lack of skills in emotional regulation may lead to poor mental health and relationship difficulties. It is as important to look after our mental health as it is to look after our physical health. For mental health professionals, the pandemic has also brought many ethical challenges. Personal protection, personal treatment needs if they get infected, impact on others if they get infected, economic crisis, ethical problems for themselves and others, and training are the issues that concern mental health professionals. In the wake of the pandemic, the training of residents has also been compromised. The ways of learning for medical students and residents may also change, leading to an opportunity to innovate. This research concentrates upon the above-mentioned purpose and tries to bring out the fact about the same. This study shows us the effect on mental health by spending more time in social media during lockdown, what its impact on their mental health during lockdown in different age groups and how to reduce spending more time on social media to avoid depression and keep mental condition positive.*

# 1. Introduction

During this lockdown people are started spending huge amount of time on social media and for many youths nowadays, social media has been popular aspects of life. Whether positive or negative, most people engage with social media without stopping to think about what the effects are on our lives. Social networking is a perfect place to rapidly distribute content across the world, with posts like "breaking news" earning hundreds of thousands of retweets in minutes. Although social media impacts people positively, it also has negative impacts on people. Social networking has been described primarily to refer to "the many fairly inexpensive and widely available online resources that make it easier for everyone to publish and access knowledge, cooperate on a collaborative project or create relationships. It has become a forum of discussion of every single social issue that is taking place. In the current situation, one of the most pathetic and to – bother social media use and increase the risk factor of mental health of humans. This has been increasing day by day at nook and corner of the country. Mindfulness simply means being in the present without thinking about the past or the future; choosing what you react to, rather than being carried away by everything that appears in your mind or experience; being non-judgmental and cultivating an attitude of impermanence towards things and situations, focusing on one thing at a time. This allows us to remain open to experiences and allows you not to be overly affected by them.

The sparse literature on the effects of epidemics on mental health relates more to the sequelae of the disease itself than to social distancing (e.g., mothers of children with congenital Zika syndrome). Large-scale disasters, however, whether traumatic (e.g., World Trade Center attacks or mass shootings), natural (e.g., hurricanes), or environmental (e.g., Deep-water Horizon oil spill), are almost always accompanied by increased depression, posttraumatic stress disorder (PTSD), drug use disorder It seems likely that there will be significant increases in anxiety and depression, substance use, loneliness, and domestic violence in the context of the COVID-19 pandemic; and with schools closed, there is a very real possibility of an epidemic of child abuse In most developed countries, suicide is the second leading cause of death among young people, exceeded only by accidents. One of the strongest known suicide predictors is self-harm. In the India both suicide and self-harm have increased in young people, and social media has played a prominent role in how young people communicate about self-harm in lockdown. No matter people are staying with their family, friends or alone reason behind this people are not allowed to go out of home and getting more free time as compare to normal situation, how long they talk with others so at last they started using Facebook, whasapp, twitter etc.

*Originality and value*: A lot of studies have been conducted an effect of social media use on mental health during lockdown human behavior but the catalytic effects of said media is still under studied. Our research aims to address this gap by undertaking a primary study to understand how social media plays the part of a mental health in lockdown. Social networking sites like Facebook and Twitter are being widely used as an open forum to discuss social and public issues like politics, human rights, climate change etc. Research on the catalytic effects of these platforms is invaluable in today's time.

Not everything is bad: the positive effects of social media on people in lockdown. Social media is here to stay and can be a tool that is useful and efficient. The following positive functions can be served by social media:

a) *Connection*: Social media, especially if you can't connect in person, can be a great way to stay in touch with friends and family. It can also help us to meet new individuals and find individuals with similar interests. Social media can also be a way to connect with individuals you admire or inspire and can be a way to inspire and motivate others.
b) *Self-promotion/Marketing*: Social media is a legitimate way to promote our brand or company and to raise awareness of your interests, hobbies, charities or business ventures. It can be a way to disseminate quality information if used effectively and can be an avenue to improve our financial position and help with future opportunities for employment.
c) *Positive feedback*: Social media can impact your psyche positively because, let's face it, praise feels good. The idea is not to post only to receive positive feedback, but if you post about something you feel good about or something you have discovered or accomplished, receiving encouragement and praise can have a positive impact on your mindset, it can improve your self-concept and it can increase your motivation to continue pursuing your goals.
d) *Health & Well-being*: In improving both their physical and psychological well-being, many individuals have found social media helpful. Examples include applications, websites, and online forums that promote self-care, resilience, and a healthy lifestyle and provide accurate information and strategies to help individuals create and maintain emotional well-being.

Social Media's Dark Side: Negative Effects of Social Media on people

a) *Self-promotion*: When you present a false view of your life for no other purpose than to get followers or to be a follower, inauthentic self-promotion is. It has no real meaning or purpose, and rather than based on a strong sense of identity and values, your self-concept is highly influenced by the perception of you by others. Instagram has recently started a trial in Australia, Brazil, Canada and four other nations, eliminating the ability for others to see how many likes a post has received, perhaps as a way to enhance the significance and authenticity of responses to posts from individuals.
b) *Vulnerable to cyber-bullying & trolling*: Participating in social media opens you up from both people you know and from strangers to threats and potentially psychologically and physically dangerous situations. The risk of being trolled or stalked is run by everyone who has set up a public profile. Alternatively, individuals, and perhaps mostly young people, have the ability to say or do something that is deemed inappropriate or irresponsible online, and this can have a significant impact on their future job prospects, their relationships, and even their community status.
c) *Isolation*: One of the most important factors in maintaining mental health and well-being is social connexion, and although social media can help increase connexions with others, it can also account for the rise in our society's loneliness. People choose to interact more and more through social media platforms and experience less and less interaction between individuals. This lack of real-life interaction practice can lead to problems such as social anxiety, avoidance behaviors, problems with self-esteem and depression.
d) *Distraction*: Social media can be a major distraction from important aspects of life, such as spending quality time or achieving life goals with loved ones. Overuse can lead to poor, procrastinating and even dependent sleeping habits. During assignments that require focus and attention, social media can make people less aware. One of the most relevant behaviours that have arisen from the rise of social media and smartphones is distraction while driving.

In this context, the purpose of our research is to find out how social media effecting and increasing the risk factor of mental health like depression during this lockdown and what will be the solution to avoid more use of social media in free time how to engage our self in other work to keep mental condition positive. In today's world, many of us rely on social media platforms to find and connect with each other, including Facebook, Twitter, Snap chat, YouTube and Instagram.

## 2. Literature Review

All social media sources play a major role in the online information ecosystem and generate engagement from millions of social media users. We define junk health news and information sources by evaluating whether or not their content is extremist, sensationalist, conspiratorial, or commentary masked as news in Lockdown. India reported the first case of COVID-19 on 30 January 2020 and the numbers have steadily increased since then, albeit at an alarming rate in the final days of March. Since 24 March 2020, the world's largest democracy has implemented the world's largest nationwide lockdown in order to control community transmission (The Lancet, 2020). Given the high population density, socio-economic fabric and overstretched health-care infrastructure, the country remains vulnerable to COVID-19. The only immediately available, best and ideal solution to the COVID-19 pandemic control in India was the overall lockdown. At multiple levels, the Indian government has responded appropriately, adequately and quickly to the COVID-19 pandemic. The lockdown has helped India to buy crucial time: time for extensive contact tracking, time to ramp up testing, and most importantly, time to prepare our health system, enhance its health care infrastructure, and prevent it from overwhelming. During this covid 19 India is under lockdown and forcing people into home confinement. We were interested in characterizing the change in social media use before going to bed, in sleep quality and timing, and in the subjective experience of time passing, and their relationship with depression, anxiety and stress levels. Some aspects of students' mental health 80 will improve (e.g., aspects of daily stress) while others will worsen (e.g., loneliness).To get a better understanding of factors explaining change in mental health, we 82 further explore the impact of different individual and social factors on change in mental 83 health. We consider COVID-19-related stressors, social network integration within the 84 student community, social ties outside the student community, and demographic factors. During crisis events, people often seek out event-related information to stay informed of what is happening. However, when information from official channels is lacking or disseminated irregularly, people may be at risk for exposure to rumors that fill the information void. Stressful life events, extended home confinement, brutal grief, interfamilial violence, overuse of the Internet and social media are factors that could influence the mental health of adolescents during this period. The COVID-19 pandemic could result in increased psychiatric disorders such as Post-Traumatic Stress, Depressive, and Anxiety Disorders, as well as grief-related symptoms. Adolescents with psychiatric disorders are at risk of a break or change in their care and management; they may experience increased symptoms. The new onset of Illness Anxiety Disorder is likely to increase the COVID-19 pandemic and country-wide lockdown and cause symptoms to worsen in diagnosed cases. Any simple flu-like sign increases anxiety and COVID-19 is expected to have a more severe effect under current circumstances. Patients with Obsessive Compulsive Disorder, particularly those with compulsive control, hoarding and washing, are at higher risk. The contamination obsessions and washing compulsions could increase advice on improving personal hygiene measures. Patients are more likely to resort to panic buying and excessive hoarding of essential items in the face of ongoing lockdown, even though the states guarantee continuous supply of essential items. For patients with recurrent depressive disorder, lockdown is a major stress that jeopardizes normal daily routine, social rhythm and thus increases levels of stress, which would further increase the level of cortisol, leading to a vicious exacerbation of depressive symptoms. This is the same for chronic insomnia, generalized anxiety disorder, and even suicide. It is the same for chronic insomnia, generalized anxiety disorder, and even suicide. Pandemics, moreover, are not only a medical phenomenon. Inability to join, diminishing finances and the long-term economic impact will have an impact on new and pre-existing common mental health disorders. At the end of the day on 24 March, India, home to about one sixth of the world's population, imposed a nationwide lockdown to combat the COVID-19 pandemic. India's lockdown continued until June 7 after multiple extensions, followed by phased relaxations, barring containment zones, where lockdowns are in place until the end of June. Although the state of Kerala in India, through aggressive testing, contact tracing, and cooked meals, exemplified the response to a pandemic of this scale, the country overall has been overwhelmed. Despite possible underreporting of COVID-19 cases and deaths, India's lockdown or stay-at-home order strategy, coupled with aggressive contact tracing, quarantining, and monitoring, is likely to have been successful in initially keeping the spread of the virus in check The existence of the Integrated Disease Surveillance Program, a national surveillance network, enabled India to deploy hundreds of public health workers in rural, suburban, and urban communities to monitor millions of citizens and identify disease clusters early enough to put in place containment measures. However, since the lockdown was announced with less than 4 hours' notice, such measures have come at a steep cost. In India, the COVID-19 outbreak began to escalate and, from 25 March 2020, a lockdown was introduced to resist disease transmission. As India is a populous country, an inevitable method of prevention was lockdown. Restrictions on different social practices and behavior were included in the decrees imposed during the lockdown. People had to spend their time confined mostly to their homes. Initially, school, college, and offices were closed and later partially or fully resumed with the help of electronic devices and Internet facilities on the virtual platform. As part of the regular schedule, extensive use of electronic equipment was included. Online classes are being carried out for school and college students, meetings as well as office work. Sleep is a behavior that includes various aspects, some of which can be evaluated using multiple quantitative parameters such as sleep latency, inertia, sleep duration, and debt while daytime sleepiness; nap details provide a qualitative overview. This change in the mode of work involving prolonged screen time is understood to affect sleep. The psychological and physiological impact of compromised sleep is well established. It can lead to mental health problems, metabolic disorders, and circadian de-synchronization. The sedentary activity involving screen exposure time is adversely affected by sleep rhythm. A non-governmental organization working on children's rights and health has already reported that 88 percent of respondents experienced a massive increase in screen exposure duration among Indian urban children. Digital media has become essential among adolescents that affect sleep duration, cause late bed time and wake time along with other sleep problems. Obesity and related co-morbidities may also result from erratic sleep behavior along with increased sedentary activities. Strengthening immunity is extremely necessary in this specific condition and good sleep practices can be helpful for this purpose. Immunity and sleep have already been observed to follow a relationship where both interact with effect. During this lockdown phase, many of our activity patterns and sleep habits had to undergo a visible transformation. The mitigation of a worldwide COVID-19 pandemic threat is crucial for human life and to reduce livelihood distortions. With considerable confirmed cases and deaths, the COVID-19 pandemic has swept into more than 200 countries and has caused public panic and mental health stress. Full lockdown with strict social distancing measures to break the transmission chain has been implemented by most nations around the world. Global health and mental health are strongly affected by the current outbreak of COVID-19. Despite all the resources used to counteract the spread of the virus, additional global strategies are needed to deal with the mental health issues involved.

People from all walks of life have been severely affected by the emergence of the COVID -19 pandemic. The rapid spread of the disease to almost all parts of the country has presented the entire human population with enormous health, economic, environmental and social challenges. The only alternatives are social distancing and other preventive measures, in the absence of any effective drugs and vaccines for treatment. Due to the complexities of pandemics and the scarcity of evidence about the epidemiology of pandemic-related MH problems and the potential interventions to address them, the emergence of mental health (MH) problems during a pandemic is extremely common, though difficult to address. Policymakers, stakeholders and researchers have so far devoted little attention to this topic, resulting in a lack of replicable, scalable and applicable frameworks to help plan, develop and deliver MH care during pandemics. As a response, during the ongoing COVID-19 pandemic, we have attempted to develop a conceptual framework (CF) that could guide the development, implementation, and assessment of MH interventions. The threat is multiplied by the additional concern of a strong alcohol lobby led by multinational corporations targeting India's emerging market for young drinkers. The company's enthusiasm is fuelled by a steady change from a culture of abstinence to ambivalence to covertly permissive in the level of acceptance and attitude towards alcohol. In order to minimize the public health impact of the aforementioned factors, India needed a national level alcohol policy. A semi-structured questionnaire was used to elicit their views. They were asked to provide their views on the current situation of MH in their countries and to elaborate on existing myths and misinformation. The course of the India Pandemic In terms of mortality and spread, there are different Infection in comparison to some other infections. The world's countries at the present time. The economic, social and psychological. The impact of the pandemic is apparent. Think that exploring the ways is essential, how individuals have found that they can deal with the Pandemic situation with social situation on one side. The other side of isolation that might have Never-before-seen-before. It is a chance to find out how people modify their routines while staying inside their homes, and habits. The concerned government, hospitals, educational institutions, organizations, and even individuals need to look into psychological intervention during such stressful situations and take the necessary measures. In addition to educating people to remain isolated, educating and preparing them to face the mental health problems they may endure during the period is essential. They were also requested to name the available resources and to suggest solution. The psychological and mental wellbeing of health care Providers are another worry. Healthcare providers constantly work in fearful, stressful, resource-constrained environments Settings where they are exposed and infected under the continuous threat of being exposed. In such a situation, the mental health and psychosocial well-being of health care providers is as important as managing the health of the infected population. There are various online resources that can be available. Helpful in managing and dealing with the stress resulting from stress about the pandemic. For individuals, it is essential to take care of their own family members and their friends. The World Health Organization (WHO) has suggested reducing the spread of the virus. On March 22, 2020,

India was quick to close its international borders and enforce the world's biggest COVID lockdown. The present study seeks to highlight, along with an analysis of lifestyle changes, the effect of the imposed national lockdown on society and the environment. The COVID-19 pandemic and lockdown may have a negative impact on the mental health of adolescents, although there is still no data on the long term impact of this crisis. Adolescents' individual, familial, and social vulnerability, as well as individual and familial coping abilities, are factors related to adolescent mental health in times of crisis.

## 3. Population Identification

### 3.1 Sampling Process

Our study focuses on the influence of social media on mental health of people during lockdown. The target population for the study is the population of India in the age group of 20-60. This population covers the demographic cohort Millennial and Generation Z, as these subsections of the populace are highly active on Social media platforms.

The data sample covers responses from most of citizens Bihar, Jharkhand and Maharashtra region who are staying in different parts of the country. It will be based on the voluntary participation of the students in the survey. Convenience sampling method will be used to gather inputs from the students. I collected the responses during July 2020. Number of responses collected are 120.

### 3.2 Data Collection

Data was collected through an online survey which was emailed to all the expected respondents. The objective of the questionnaire prepared was to find out the respondents' engagement to the social platforms as well as the traditional mediums i.e. electronic or print media and it intended to find out the impact of the media to the respondent.

**Primary Data:**

Main source of primary data has included the responses from students who are pursuing higher education in Pune as well as some of our friends and families. The demographic data of the respondents was collected with appropriate privacy in scales like Liker and Agree- Disagree form. The primary data includes the usage of all social media platforms like Facebook, Instagram etc. and we tried to inculcate the people aged between 20-60 in order to determine different correlates that impact the mental health of people during lockdown.

### 3.3 Objectives

1. To study whether use of social media increases depression, irritation and suicide rate in people during this lockdown.
2. To study whether people are spending most of their free time in social and how it is going to impact mental health of different age group.
3. To study if one particular social media platform has more impact on the mental health of people when compared to others.
4. Social media platforms like Facebook, Instagram etc. and I tried to inculcate the people aged between 20-60 in order to determine different correlates that effect of social media use on mental health during lockdown.

**3.4 Statistical Tools**

To perform the analysis of the collected data we used the tools like Microsoft Excel and Tableau.

**Assumptions**

We will be presuming that the Convenience sampling should be taken into consideration. The sample size of the data will be sufficiently large. The sample correctly represents the population. It is assumed that the personal issues of the respondents have no impact on their mental health and nature in regards to social media users in this situation.

## 4. Methodology

To Carry out the analysis I have followed the below methodology

1. Questionnaire has prepared to know view of people related to this topic.
2. Data was collected through online survey where 14 question was add to know view or mind set of people during lockdown.
3. Almost 120 response I got from this survey to performed analysis I use satirical tool Microsoft Excel.
4. I have found the how many people is using social media and what is the purpose of using like for news, entertainment, social communication or for some other reason.
5. I have found no of people they belongs to which occupation and how much time they are spending on social media during this lockdown daily.
6. I have found no of people how much time they spend on social media and which time they use to sleep at night.
7. I have found how many people if they are depress, upset or sad while communicating in social media with their friends or family helped them to overcome from that situation.

## 5. Analysis and Result

Data is collected from people who use social media to analyze the mental health of humans. The people who use social media and who do not use social media were interviewed on the same set of questions regarding the way they of using social media in free time during lockdown. Altogether I have frames five types of reactions a person could react in such scenarios. Respondents were asked to answer those questions. Then we calculated the average reactive score of people based on these questions of the questionnaire. This average reactive score is considered as the quantity which denotes the reaction of a person towards social media users.

*1) No of people vs Purpose of using social media*

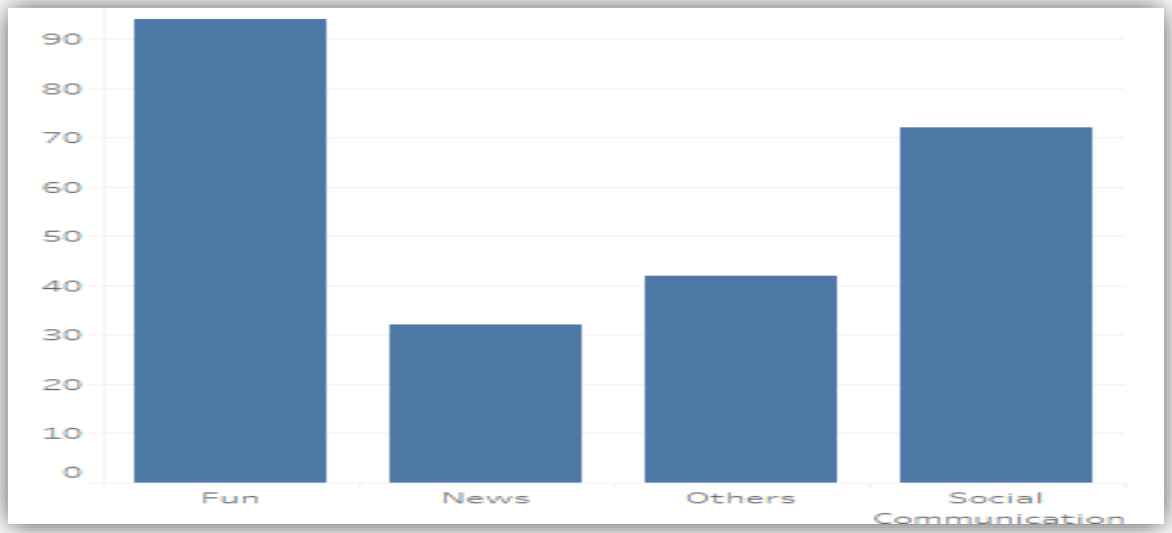

Figure 1: Purpose of using social media.

Number of people using social media, not for a platform for communication and sharing data I have observed from data most of the people is using social media for Fun/Entertainment is most those are watching movies on YouTube, watching Facebook video, posting some important thing of life on instagram, using instagram for live session. In these day people like to spend much time on social media. Social networking is a better for TV and movies than physical content. The global dialog about events and interactions viewed by people based on what they see tells us about customer preferences. Most specifically, conduct is influenced by their action.

*2) No of People Vs Occupation/Spend time on social media*

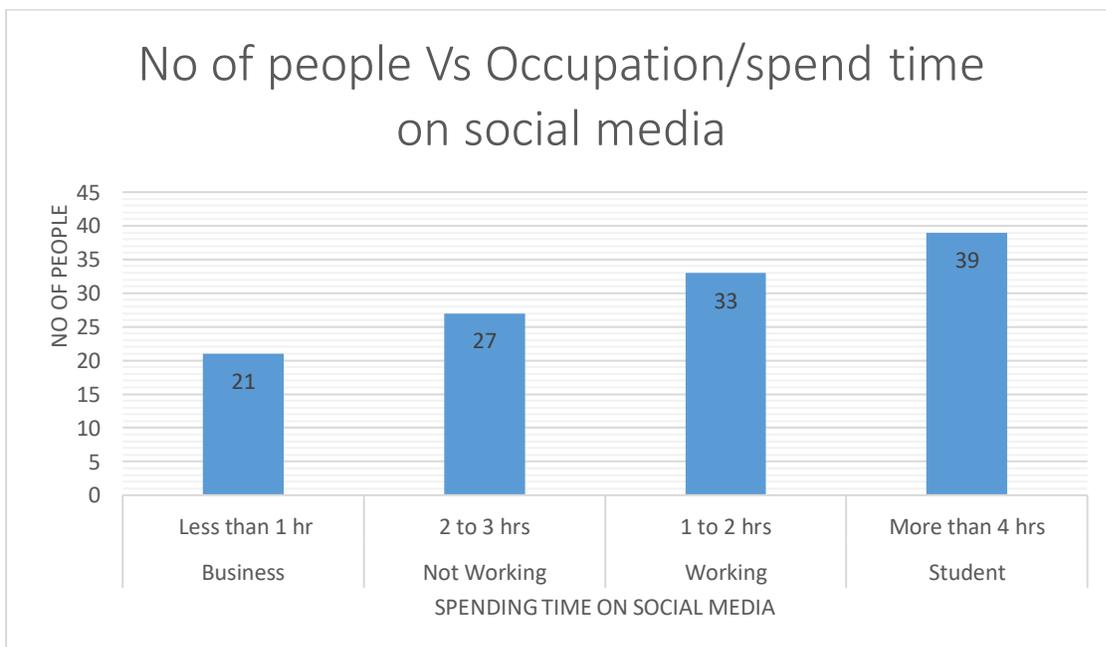

Figure 2: Occupation and spend time on social media.

Student use to spend more than 4 hrs. Time on social media daily because it's convenient for all media to communicate with people who have more control than you do. If they have more Instagram followers, own a business, play in a sweet band or have a public office, social media makes connections to celebrities from all levels easier than ever. They can check various views with social media and see which ideas we prefer. You have the freedom to share your opinions and should change them as much as possible. There are clear reasons why people use social media for hours.

*3) No of people Vs Sleep time/Spend time on social media*

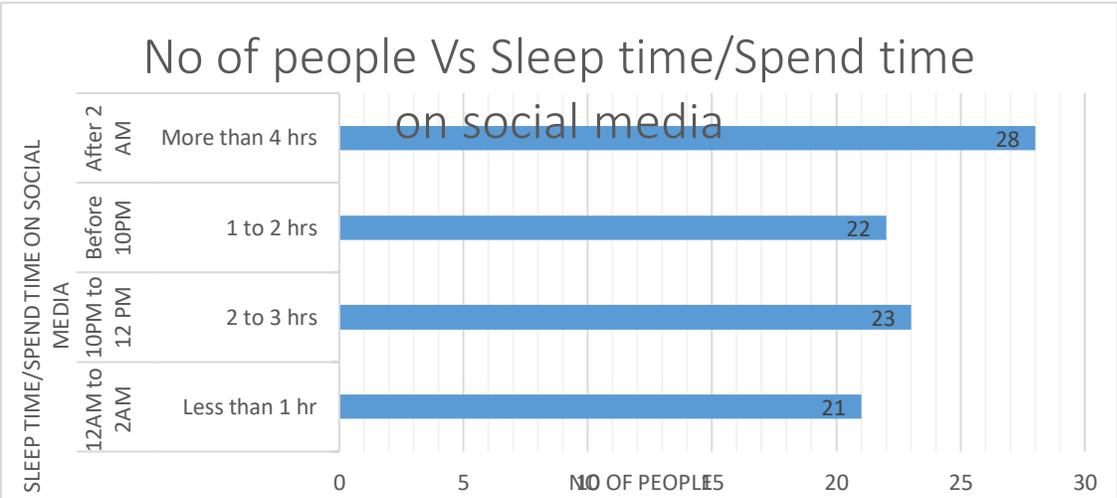

Figure 3: Spend time on social media.

Social media usage around bedtime can negatively effect how long and how well you sleep. Looking at social media in bed can make it harder for you to fall asleep. It can also reduce the amount of time you sleep for and leave you feeling unrefreshed the next day. I have observed from survey people who is spending late night on social media they sleep late night which is not good for mental health.

*4) No of People Vs person over social media affecting positively to you while in depression*

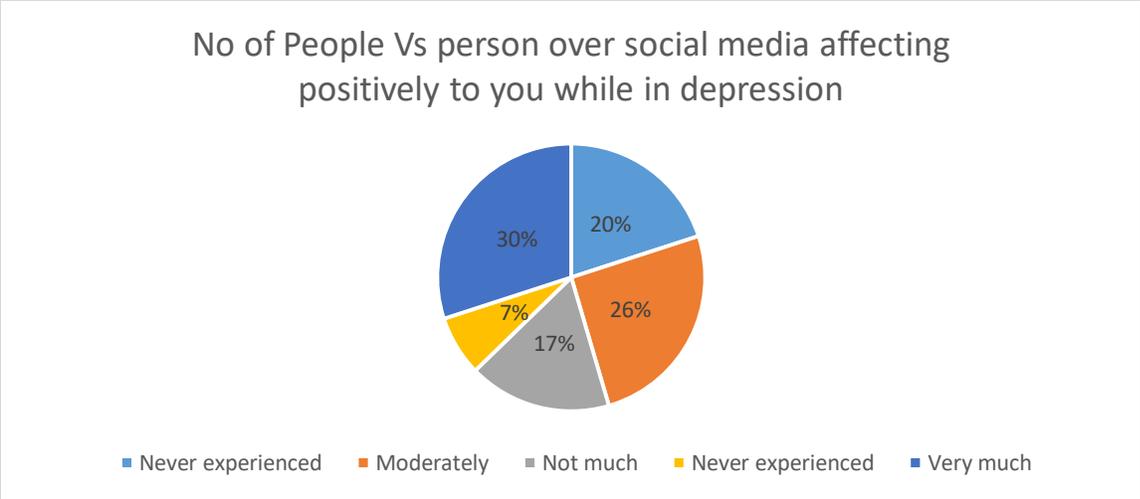

Figure 4: Social Media affection positively in depression.

Many people enjoying staying in social media in my survey I found while in depression social media is useful for people very much because its connect people if someone is in tension or depression they can communicate with their family, parents or friends.

*5) No of People Vs Activity for peace of mind*

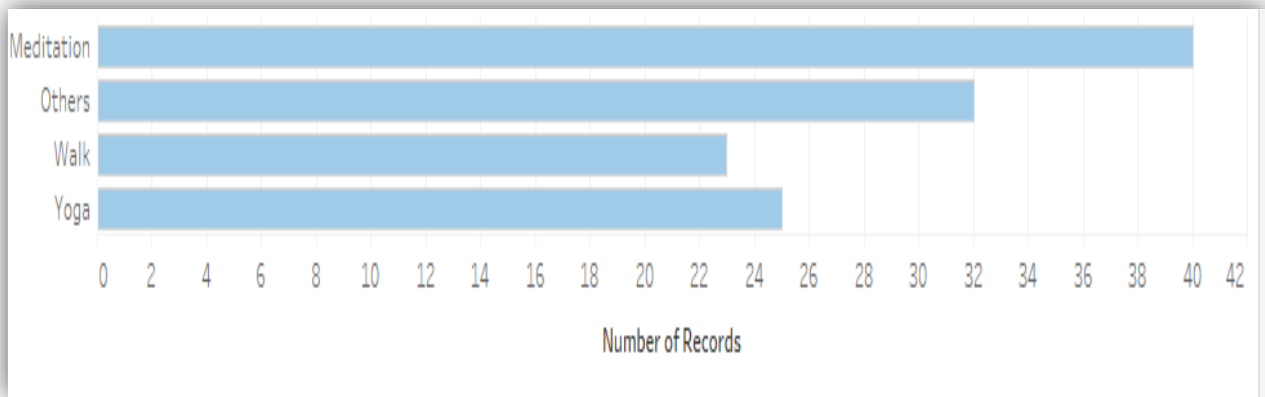

Figure 5: Activity for peace of mind

Most of people do meditation to keep mind peace during lockdown, Meditation practice also helps in knowing our bodies. We sense physical movements when we become more sensitive and aware. This is not easy, it takes time and regular practice but gives relative to state of body mental and physical.

*6) Rate of Social Media to reduce the Stress during lockdown.*

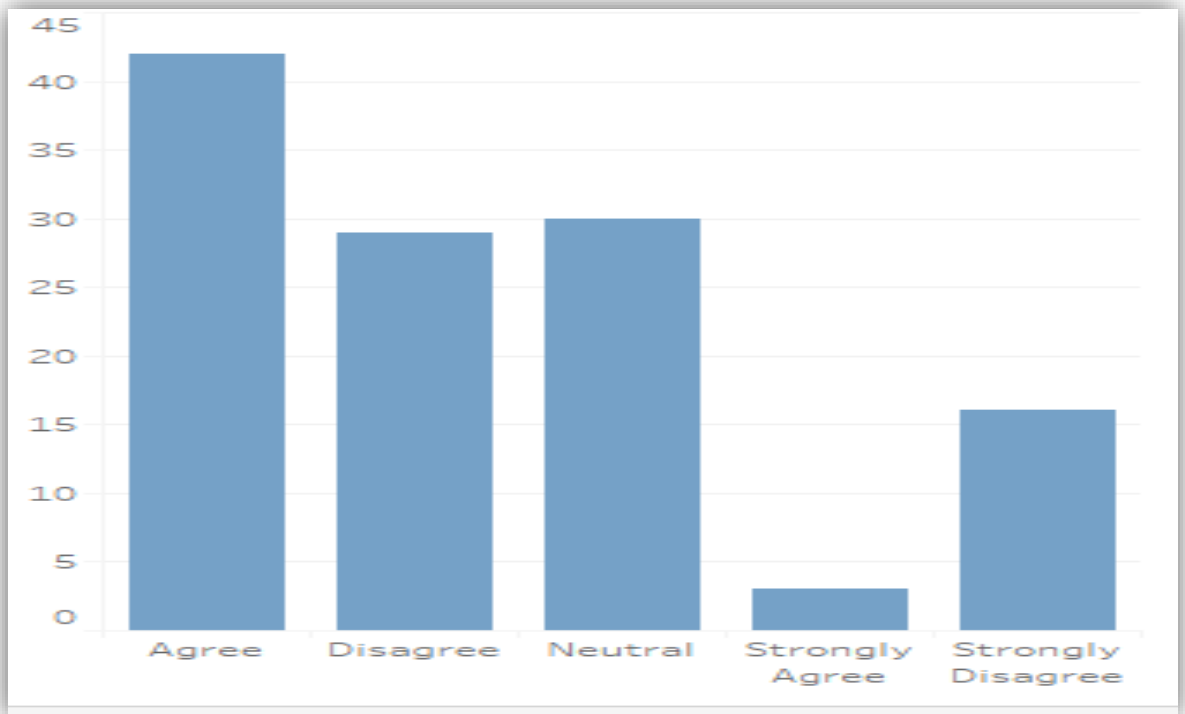

Figure 6:Feedback of reducing stress by social media during lockdown

Maximum people are agree that social media use reduce the stress during lockdown no one can meet their family or friends but they can communicate with them through social media.

## 6. Conclusion

I have found through this research most of the people use social media for their entertainment purpose during covid time. Student use to spend much time in social media all most 3 to 4 hrs. Per day to spend their time. Communicating in social media with friends and family is helping most of the people to overcome from depression, Irritation or any negativity. People like to do yoga to keep them positive or mentally stable. Most of working people feeling happy in lockdown because they got time to spend with their family but student are feeling irritated or bore because class in not continue they do not have much work to do because of this they use most of the time social media. People does mediation to keep mental health balance during lockdown mediation gives relief to physical and mental health both. Suggestion given by most people to be positive in this situation is do what you love to do. It may be writing, reading, watching movies, if you are at home then help your parents with work, talk to them spend time with them, talk to your friends via phone call or video call. This may help people to avoid negativity, depression or any mental health issue.

## 7. Limitations

My sample contains majority of people of age group between 20 - 40 years. Hence whatever conclusions are made cannot be inferred to all age groups.

All most 80 % response got from male gender so we cannot inferred to all gender whatever conclusions are made.

Also, the sample contains people mostly from Pune and whatsoever conclusion would be made cannot be inferred to the world population.

I collected 120 responses. So, the results cannot be generalized due to the small sample size.

The questionnaire and research was carried only for a short period of time. Hence, the conclusion might not be true in the future.

Lastly, not all the people might be honest while answering the questionnaire. So, the validation of the responses is a question.

# References

[1] Holmes, E.A.; O'Connor, R.C.; Perry, V.H.; Tracey, I.; Wessely, S.; Arseneault, L.; Ballard, C.; Christensen, H.; Cohen Silver, R.; Everall, I.; et al. Multidisciplinary research priorities for the COVID-19 pandemic: A call for action for mental health science. Lancet Psychiatry 2020, 7, 547–560. [CrossRef]

[2] Brooks,S.K.;Webster,R.K.;Smith,L.E.;Woodland,L.;Wessely,S.;Greenberg,N.;Rubin,G.J. Thepsychological impactofquarantineandhowtoreduceit: Rapidreviewoftheevidence. Lancet2020, 395, 912–920. [CrossRef]

[3] Taylor, M.R.; Agho, K.E.; Stevens, G.J.; Raphael, B. Factors influencing psychological distress during a disease epidemic: Data from Australia's first outbreak of equine influenza. BMC Public Health 2008, 8, 347. [CrossRef]

[4] Hawryluck, L.; Gold, W.L.; Robinson, S.; Pogorski, S.; Galea, S.; Styra, R. SARS control and psychological effects of quarantine, Toronto, Canada. Emerg. Infect. Dis. 2004, 10, 1206–1212. [CrossRef]

[5] Altena, E.; Baglioni, C.; Espie, C.A.; Ellis, J.; Gavriloff, D.; Holzinger, B.; Schlarb, A.; Frase, L.; Jernelöv, S.; Riemann, D. Dealing with sleep problems during home confinement due to the COVID-19 outbreak: Practical recommendations from a task force of the European CBT-I Academy. J. Sleep Res. 2020, e13052. [CrossRef]

[6] Ahmed, M.Z.; Ahmed, O.; Aibao, Z.; Hanbin, S.; Siyu, L.; Ahmad, A. Epidemic of COVID-19 in China and associated Psychological Problems. Asian J. Psychiatry 2020, 51, 102092. [CrossRef]

[7] Huang, Y.; Zhao, N. Generalized anxiety disorder, depressive symptoms and sleep quality during COVID-19 outbreak in China: A web-based cross-sectional survey. Psychiatry Res. 2020, 288, 112954. [CrossRef]

[8] Lei, L.; Huang, X.; Zhang, S.; Yang, J.; Yang, L.; Xu, M. Comparison of prevalence and associated factors of anxiety and depression among people affected by versus people unaffected by quarantine during the COVID-19 epidemic in Southwestern China. Med. Sci. Monit. 2020, 26, e924609. [CrossRef]

[9] Zhang, C.; Yang, L.; Liu, S.; Ma, S.; Wang, Y.; Cai, Z.; Du, H.; Li, R.; Kang, L.; Su, M.; et al. Survey of insomnia and related social psychological factors among medical staff

involved in the 2019 novel coronavirus disease outbreak. Front. Psychiatry 2020, 11, 306. [CrossRef]

[10] Yuan, S.; Liao, Z.; Huang, H.; Jiang, B.; Zhang, X.; Wang, Y.; Zhao, M. Comparison of the indicators of psychological stress in the population of Hubei Province and non-endemic provinces in China during two weeks during the coronavirus disease 2019 (COVID-19) outbreak in February 2020. Med. Sci. Monit. 2020, 26, e923767. [CrossRef]

[11] Government of Italy Decree of the President of the Council of Ministers 9 March 2020. Available online: https://www.gazzettaufficiale.it/eli/id/2020/03/09/20A01558/sg (accessed on 30 May 2020).

[12] González-Sanguino, C.; Ausín, B.; Castellanos, M.Á.; Saiz, J.; López-Gómez, A.; Ugidos, C.; Muñoz, M. Mental health consequences during the initial stage of the 2020 Coronavirus pandemic (COVID-19) in Spain. Brain Behav. Immun. 2020. [CrossRef]

[13] Ozamiz-Etxebarria, N.; Dosil-Santamaria, M.; Picaza-Gorrochategui, M.; Idoiaga-Mondragon, N. Stress, anxiety, and depression levels in the initial stage of the COVID-19 outbreak in a population sample in the northern Spain. Cad. Saude Publica 2020, 36, e00054020. [CrossRef] 2

[14] Bert, F.; Lo Moro, G.; Corradi, A.; Acampora, A.; Agodi, A.; Brunelli, L.; Chironna, M.; Cocchio, S.; Cofini, V.; D'Errico, M.M.; et al. Prevalence of depressive symptoms among Italian medical students: The multicentre cross-sectional "PRIMES" study. PLoS ONE 2020, 15, e0231845. [CrossRef]

[15] Ahmad, A.R.; Murad, H.R. The impact of social media on panic during the COVID-19 pandemic in Iraqi Kurdistan: Online questionnaire study. J. Med. Internet Res. 2020, 22, e19556. [CrossRef].

[16] Garfin, D.R.; Silver, R.C.; Holman, E.A. The novel coronavirus (COVID-2019) outbreak: Amplification of public health consequences by media exposure. Health Psychol. 2020, 39, 355–357. [CrossRef].

[17] Newman M, Zainal N. The value of maintaining social connections for mental health in older people. Lancet Public Health; Jan 2020 retrieved from https://www.thelancet.com/journals/lanpub/ article/PIIS2468-2667(19)30253-1/fulltext Richard Armitage March 19, 2020 COVID.

[18] IANS, April 11, 2020, Work From Home, Amid Covid-19 Lockdown, Changed Sleep Schedule Of 67% Indians: Survey; https://www.news18.com/news/lifestyle/wor k-from-home-amid-covid-19-lockdown- changed-sleep-schedule-of-67-indians- survey-2573703.html

[19] Nass SJ, Levit LA, Gostin LO. Beyond the HIPAA privacy rule: enhancing privacy, improving health through research: National Academies Press; 2009.

[20] Misro A, Hussain M, Jones T, Baxter M, Khanduja V. A quick guide to survey research. The Annals of the Royal College of Surgeons of England. 2014; 96(1):87

[21] 2019 Novel coronavirus, Wuhan, China. 2020. https://www.cdc.gov/ coronavirus/2019-nCoV/summary.html. Accessed 1 Feb 2020. 16. National Health Commission of People's Republic of China. Notice on printing and distributing the work plan for prevention and control of pneumonia caused by novel coronavirus infection in the near future. 2020.

[22] World Health Organization. (2020b). Helping children cope with stress during the 2019-nCoV outbreak. https://www.who.int/docs/default- source/coronaviruses/helping-children-cope-with-stress-print.pdf? Sfvrsn=f3a063ff_2&ua=1 World Health Organization. (2020c).

[23] Mental health and COVID-19. http://www.euro.who.int/en/health-topics/health-emergencies/cor- onavirus-covid-19/novel-coronavirus-2019-ncov-technical-guidance/coronavirus-disease-covid-19-outbreak-technical-guidance-europe/ mental-health-and-covid-19

[24] World Health Organization. (2020d). Mental health and psychosocial considerations during the COVID-19 outbreak. WHO reference num- ber: WHO/2019-nCoV/Mental Health/2020.1? https://www.who.int/ docs/default-source/coronaviruse/mental-health-considerations.pdf.

[25] Hall, R., Hall, R., & Chapman, M. (2008). The 1995 Kikwit Ebola outbreak: Lessons hospitals and physicians can apply to future viral epidemics. General Hospital Psychiatry, 30(5), 446–452. Https: //doi.org/10.1016/j.genho sppsy ch.2008.05.003.

[26] Smith, R.D. Responding to global infectious disease outbreaks: Lessons from SARS on the role of risk perception, communication and management. Soc. Sci. Med. 2006, 63, 3113–3123. [CrossRef]

[27] Ng, E. The pandemic of hate is giving novel coronavirus disease (COVID-19) a helping hand. Am. J. Trop. Med. Hyg. 2020, 102, 1158–1159. [CrossRef]

[28] Tonkin son, P. How Did Runners Become Public Enemy Number One? The Telegraph. Available online: https: //www.telegraph.co.uk/health-fitness/body/did-runners-become-public-enemy-number-one/ (accessed on 30 May 2020).

[29] Jónsdóttir, U.; Þórðardóttir, E.B.; Aspelund, T.; Jónmundsson, Þ.; Einarsdóttir, K. The effect of the 2008 recession on well-being and employment status of people with and without mental health problems. Eur. J. Public Health 2019. [CrossRef]

[30] ISTAT Istituto Nazionale di Statistical. Indicatori Demografici Anno 2019; ISTAT Istituto Nazionale di Statistica: Rome, Italy, 2020.